\documentclass[twocolumn,showpacs,preprintnumbers,amsmath,amssymb]{revtex4}
\usepackage{graphicx}
\begin{document}
\title{Spin dependent conductance of a quantum dot side attached to topological superconductors as a probe of Majorana fermion states}
\author{D. Krychowski$^{1}$}
\email{krychowski@ifmpan.poznan.pl}
\author{S. Lipi\'{n}ski$^{1}$}
\author{G. Cuniberti$^{2}$}
\affiliation{%
$^{1}$Institute of Molecular Physics, Polish Academy of Sciences, Smoluchowskiego 17,
60-179 Pozna\'{n}, Poland}
\affiliation{%
$^{2}$Institute for Materials Science and Max Bergmann Center of Biomaterials, TU Dresden, 01062 Dresden, Germany}
\date{\today}
\begin{abstract}
Spin-polarized transport through a quantum dot side attached to a topological superconductor  and coupled to a pair of normal leads is discussed in Coulomb and Kondo regimes.  For discussion of Coulomb range equation of motion method with  extended Hubbard I approximation is used and  Kondo regime is   analyzed  by Kotliar-Ruckenstein slave boson approach.  Apart from the occurrence of zero bias anomaly  the presence of   Majorana states reflects also  in  splitting of Coulomb lines.  In the region of Coulomb borders   the spin dependent negative differential conductance is observed. Due to the low energy scale of Kondo effect   this probe allows for detection  of Majorana states  even for  extremely weak coupling with  topological wire. In this range  no signatures  of Majorana states  appear in  Coulomb blockade dominated transport.
\end{abstract}

\maketitle

\section{Introduction}
Majorana fermions and Majorana bound states (MBSs) have attracted considerable interest in recent years due to their fundamental exotic properties e.g. self-hermicity and  associated with this peculiarity  non-Abelian statistics \cite{Leijnse}.  These features make Majorana states  potential candidates for the use in fault-tolerant topological quantum computation \cite{Sarma}.  MBSs are predicted to exist at the ends of a semiconductor nanowire with strong spin-orbit coupling placed  in external magnetic field and  brought into proximity of s-wave superconductor \cite{Mourik}. Also several other propositions for  the  realizations of 1d topological superconductor (TS) have been reported e.g.   carbon nanotubes with broken chiral symmetry and   curvature induced spin-orbit coupling \cite{Sau}. Various proposals have been made to detect the Majorana states using different hybrid structures based on quantum dots \cite{Lee,Stefanski}. In this report we discuss signatures of Majorana states in transport through quantum dot coupled to normal leads and to a single or a pair of TS wires.  Both Coulomb blockade range and Kondo regime are discussed.  Due to  the helical properties of TS wire its Majorana end-state hybridizes with only one of the dot spin orientations \cite{Leijnse,Sticlet}, and thus in addition to conductance also polarization of conductance gives information on Majorana states. Present report complements the  earlier studies on this topic by detailed analysis of spin polarization. We present the possibility of control of spin transport and we find the spin negative differential conductance, the phenomenon of wide array of potential applications e.g. in spin dependent amplifiers or switching circuits.

\section{Model and formalism}
T-shape system of Majorana bound state coupled to the quantum dot is presented in Fig. 1a (single side attached TS) and on the inset of Fig. 2a (a pair of TSs).  The total Hamiltonian is  ${\cal{H}} = {\cal{H}}_{0}+{\cal{H}}_{MBS}+{\cal{H}}_{DM}$.  ${\cal{H}}_{0}$ is the Anderson hamiltonian and it describes the dot and the leads:
 \begin{eqnarray}
&&{\mathcal{H}}_{0}=\sum_{\sigma}E_{d}n_{\sigma}+{\cal{U}}n_{\uparrow}n_{\downarrow}\nonumber\\&&+\sum_{k\alpha=L,R \sigma}\varepsilon_{k\alpha \sigma}n_{k\alpha \sigma}+\sum_{k\alpha \sigma}(Vc^{\dagger}_{k\alpha \sigma}d_{\sigma}+h.c.)
\end{eqnarray}
where $E_{d}$ is QD level and  $\varepsilon_{k\alpha\sigma}$  are energies of conduction electrons. The term parameterized by ${\cal{U}}$ describes Coulomb interaction and $V$ represents hopping between the dot and normal leads.
${\cal{H}}_{MBS}=\sum_{l}i\delta_{l}\gamma^{l}_{1}\gamma^{l}_{2}$ describes  coupling between the two Majorana states $\gamma^{l}_{1}$, $\gamma^{l}_{2}$  in the $l$-th  TS wire ($l = 1,2$). Coupling between Majorana modes can be neglected   in the case when TS wire is much longer than the coherence length ($L\gg\xi$). In the opposite limit (dirty topological superconductors)  $\delta\neq0$ ($\delta\sim e^{-L/\xi}$). The  Majorana operators can be expressed by fermionic operators:  $\gamma^{l}_{1}=(f_{l}+f^{\dag}_{l})/\sqrt{2}$, $\gamma^{l}_{2}=i(f_{l}-f^{\dag}_{l})/\sqrt{2}$. Due to helical properties of TS wire Majorana states are spin polarized \cite{Sticlet}. The dot-Majorana coupling term is ${\cal{H}}_{DM}=t_{1} \gamma^{1}_{1}(-d_{\uparrow}+d^{\dagger}_{\uparrow})+it_{2} \gamma^{2}_{2}(d_{\downarrow}+d^{\dagger}_{\downarrow})$, where we assumed configuration in which spin polarizations of Majorana states  $\gamma^{1}_{1}$   and  $\gamma^{2}_{2}$ are opposite and thus  electrons of one spin direction, say up,  are coupled with MBS from the lower wire and spin down electrons with Majorana state of the upper wire.
To find QD  retarded Green's functions $G^{R}(t)=\langle \langle d_{\sigma}(t);d^{\dag}_{\sigma}(0)\rangle \rangle$ in Coulomb blockade regime we use extended Hubbard I approximation. In Hubbard I approach two-particle Green's function $\langle \langle d_{\uparrow}n_{\downarrow};d^{\dag}_{\uparrow}\rangle \rangle$ are decoupled as follows $\langle n_{\downarrow}\rangle\langle \langle d_{\uparrow};d^{\dag}_{\uparrow}\rangle \rangle$, and in the extension we use they are treated exactly.
\begin{figure}
\includegraphics[width=0.43\linewidth]{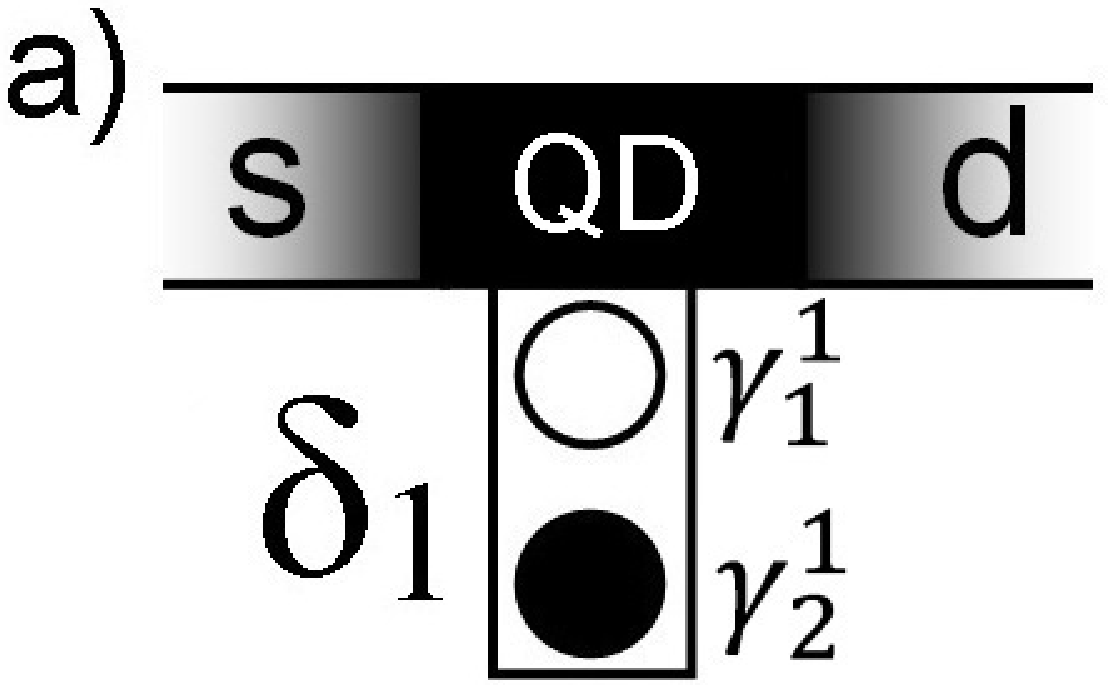}
\includegraphics[width=0.48\linewidth]{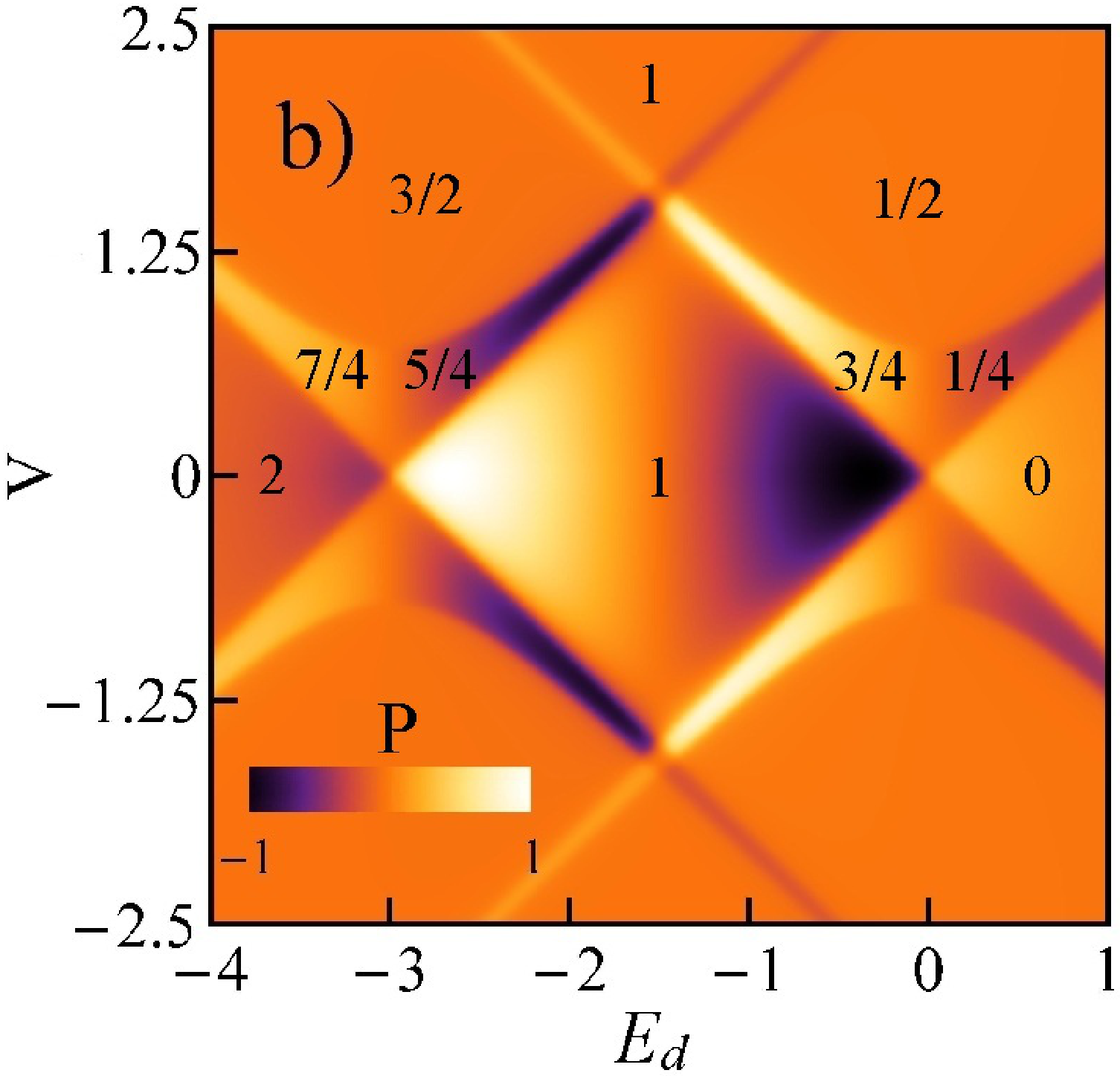}\\
\includegraphics[width=0.48\linewidth]{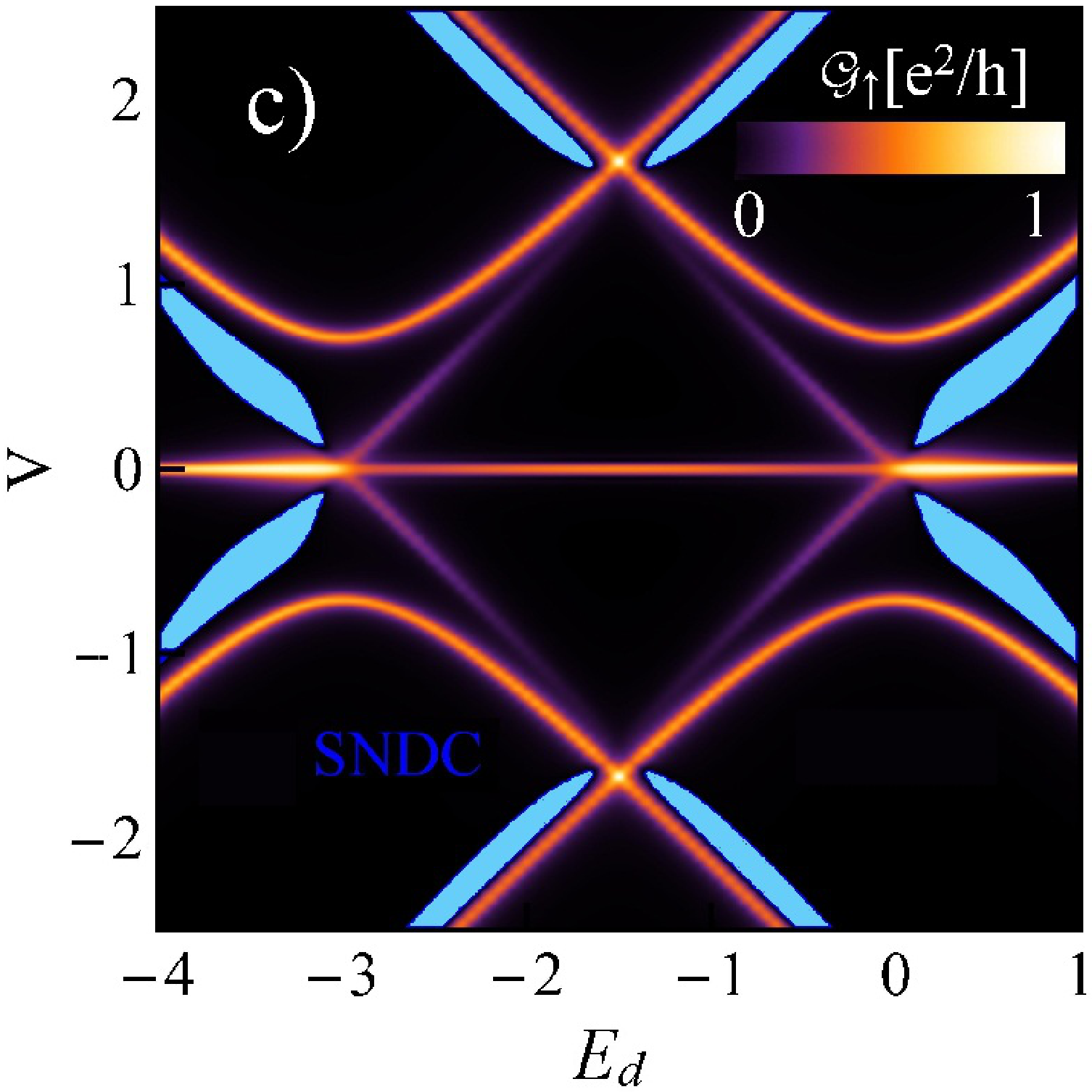}
\includegraphics[width=0.48\linewidth]{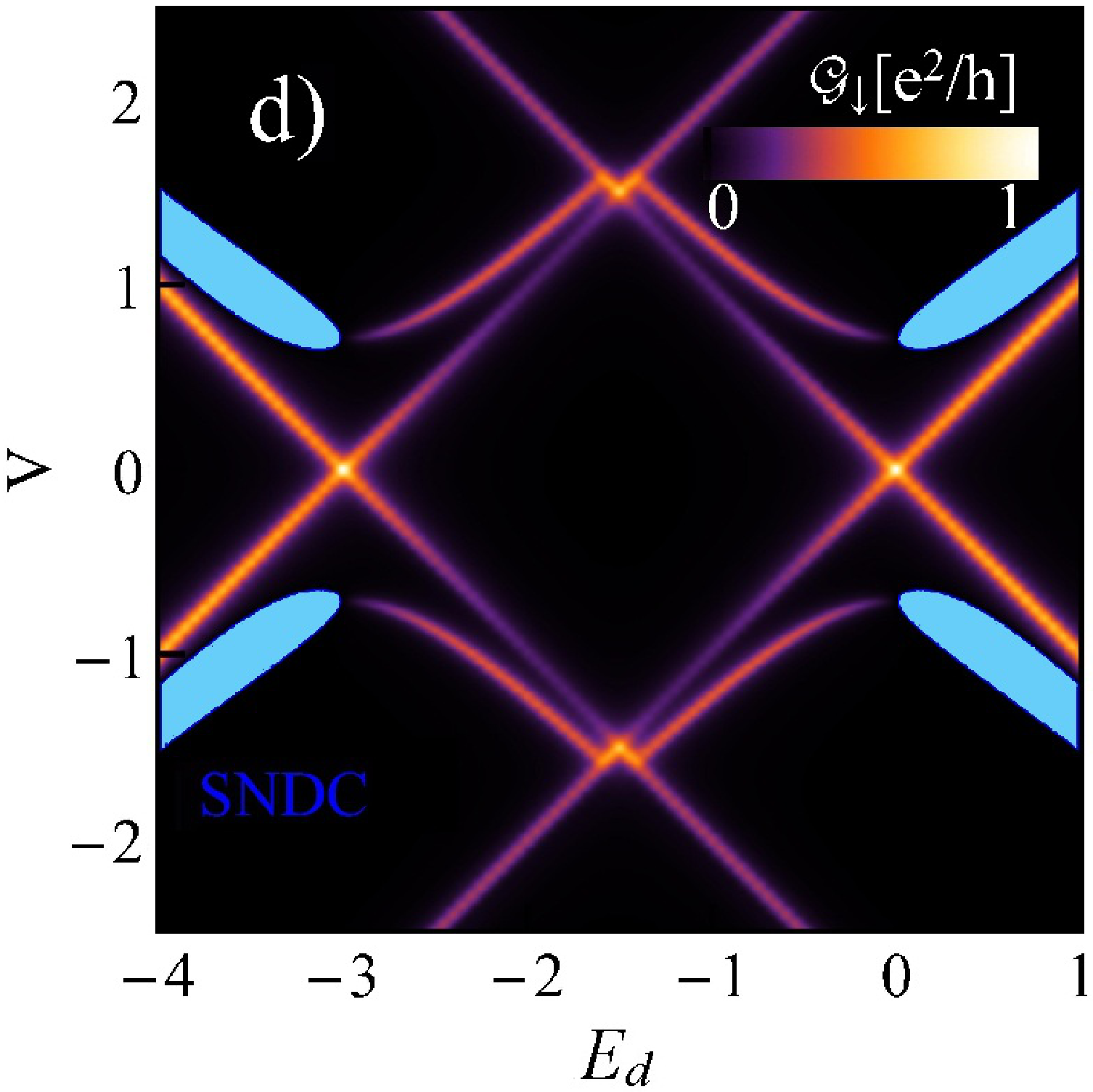}\\
\includegraphics[width=0.48\linewidth]{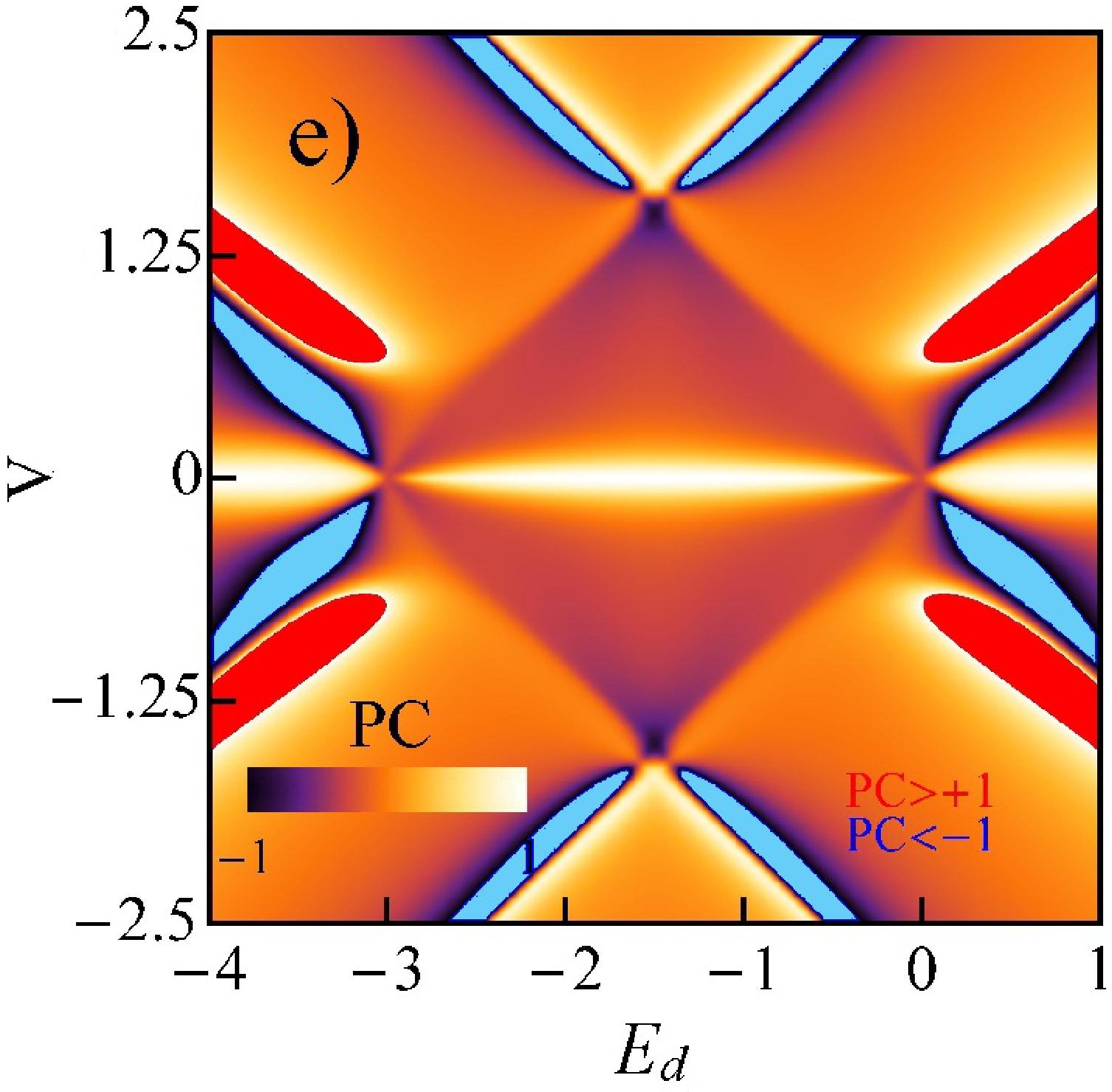}
\includegraphics[width=0.48\linewidth]{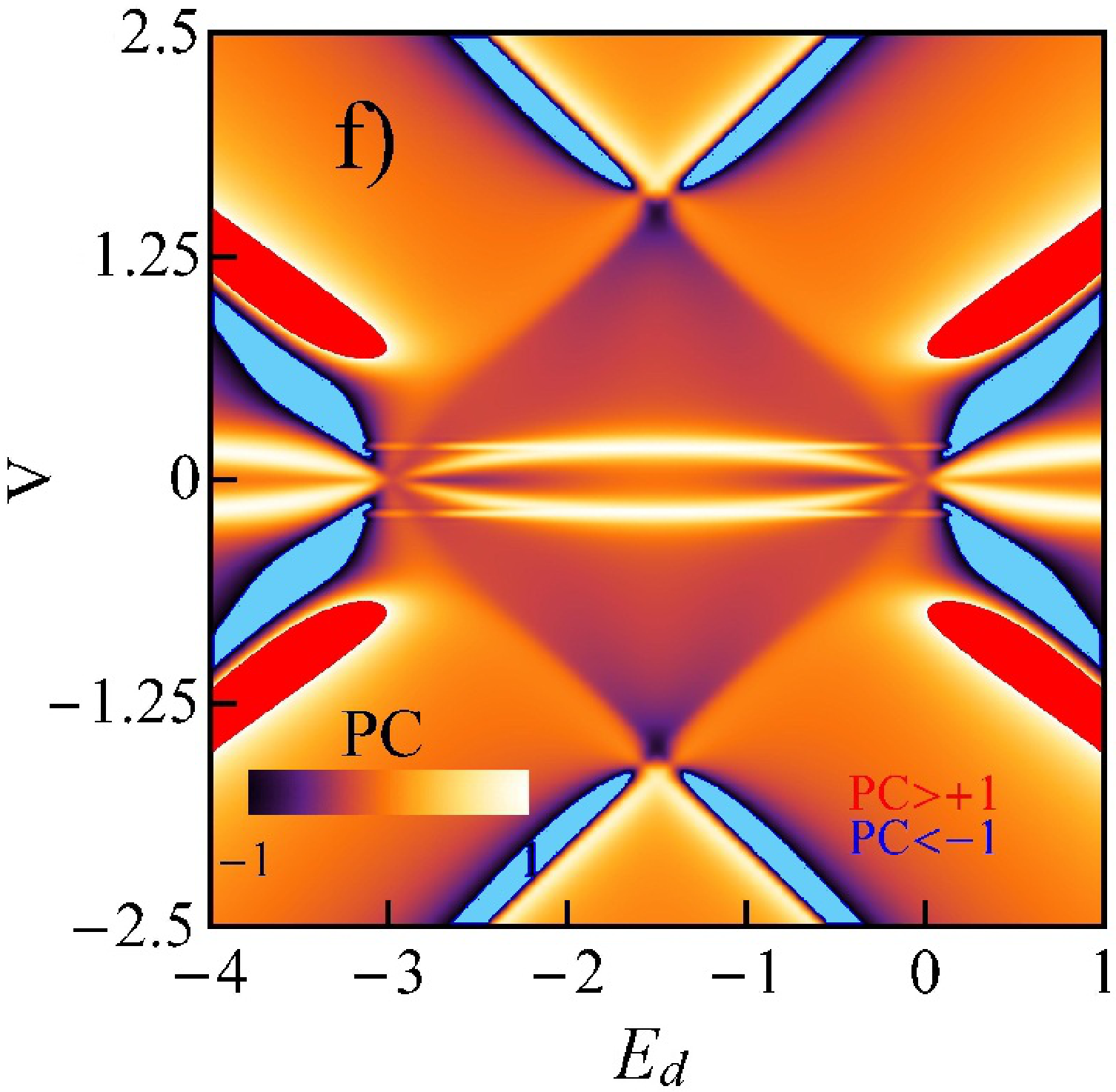}
\caption{\label{fig1} (Color online) a) Scheme of a quantum dot coupled to topological superconducting wire in T-shape geometry. White and black circles symbolize two Majorana states of opposite spin polarizations. Coulomb blockade regime: b) Spin accumulation map with occupation numbers on it. c) Conductance map for spin up orientation ${\cal{G}}^{\uparrow}(E_{d},V)$ (blue areas denote the  SNDC regions  for this spin orientation.  d)  ${\cal{G}}^{\downarrow}(E_{d},V)$  with marked SNDC regions for spin down orientation. e,f ) Plots of polarizations of conductance for $\delta =0$ (e) and $\delta =0.2$ (f). Red and blue regions correspond to $PC>+1$ and $PC<-1$ respectively (${\cal{U}} = 3$, $\Gamma = 0.05$, $t = 0.5$, the energy unit is defined in terms of the bandwidths $2D=100$).}
\end{figure}
Kondo regime is discussed within finite $\cal{U}$ slave boson approach (K-R)\cite{Kotliar}, which maps the problem into the effective noninteracting particles picture with  renormalized dot energy, hoppings to the leads and coupling to Majorana state.
Since the only term not discussed so far in K-R formalism is ${\cal{H}}_{DM}$, we give here its SB representation
${\cal{H}}^{SB}_{DM}=t_{1} \gamma^{1}_{1}(z^{\dag}_{\uparrow}f^{\dagger}_{\uparrow}-z_{\uparrow}f_{\uparrow})+i t_{2} \gamma^{2}_{2}(z^{\dag}_{\downarrow}f^{\dagger}_{\downarrow}+z_{\downarrow}f_{\downarrow})$, where  $z_{\sigma}=(e^{\dag}p_{\sigma}+p^{\dag}_{\overline{\sigma}}d)/\sqrt{\langle Q_{\sigma}\rangle(1-\langle Q_{\sigma}\rangle)}$  is the renormalization factor, with SB operator $e$ acting as projector onto empty state, $p_{\sigma}$ onto single occupied state and $d$ onto doubly occupied state. $Q_{\sigma}=p^{\dag}_{\sigma}p_{\sigma}+d^{\dag}d$ is SB representation of the occupation number operator.
Current flowing through the dot is expressed by  ${\cal{I}}_{\sigma}=(2e/h)\int^{+\infty}_{-\infty}(f_{L}-f_{R})T_{\sigma}(E)dE$, where $f_{L(R)}$ are the Fermi-distribution functions in the left and right electrodes and transmission  $T_{\sigma}=-\Gamma_{\sigma}\Im[G^{R}_{\sigma,\sigma}]$. $\Gamma_{\sigma}=2\pi V^{2}\varrho_{\sigma}(E_{F})$, and $\varrho$ is DOS of the leads.
The differential conductances are given by ${\cal{G}}_{s}=d{\cal{I}}_{s}/dV$ and
the spin polarization of conductance is defined by $PC=({\cal{G}}_{\uparrow}-{\cal{G}}_{\downarrow})/{\cal{G}}$.
\begin{figure}
\includegraphics[width=0.48\linewidth]{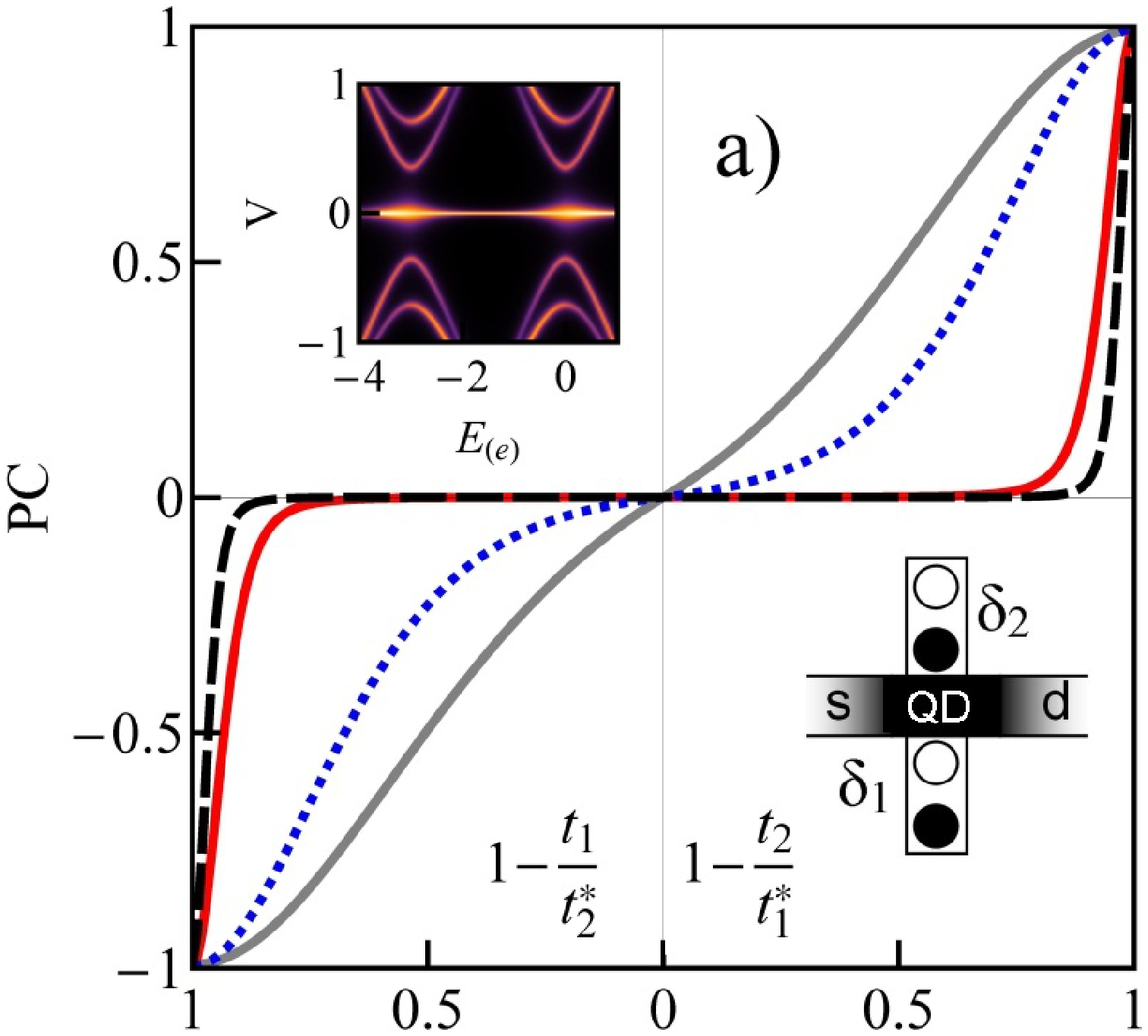}
\includegraphics[width=0.48\linewidth]{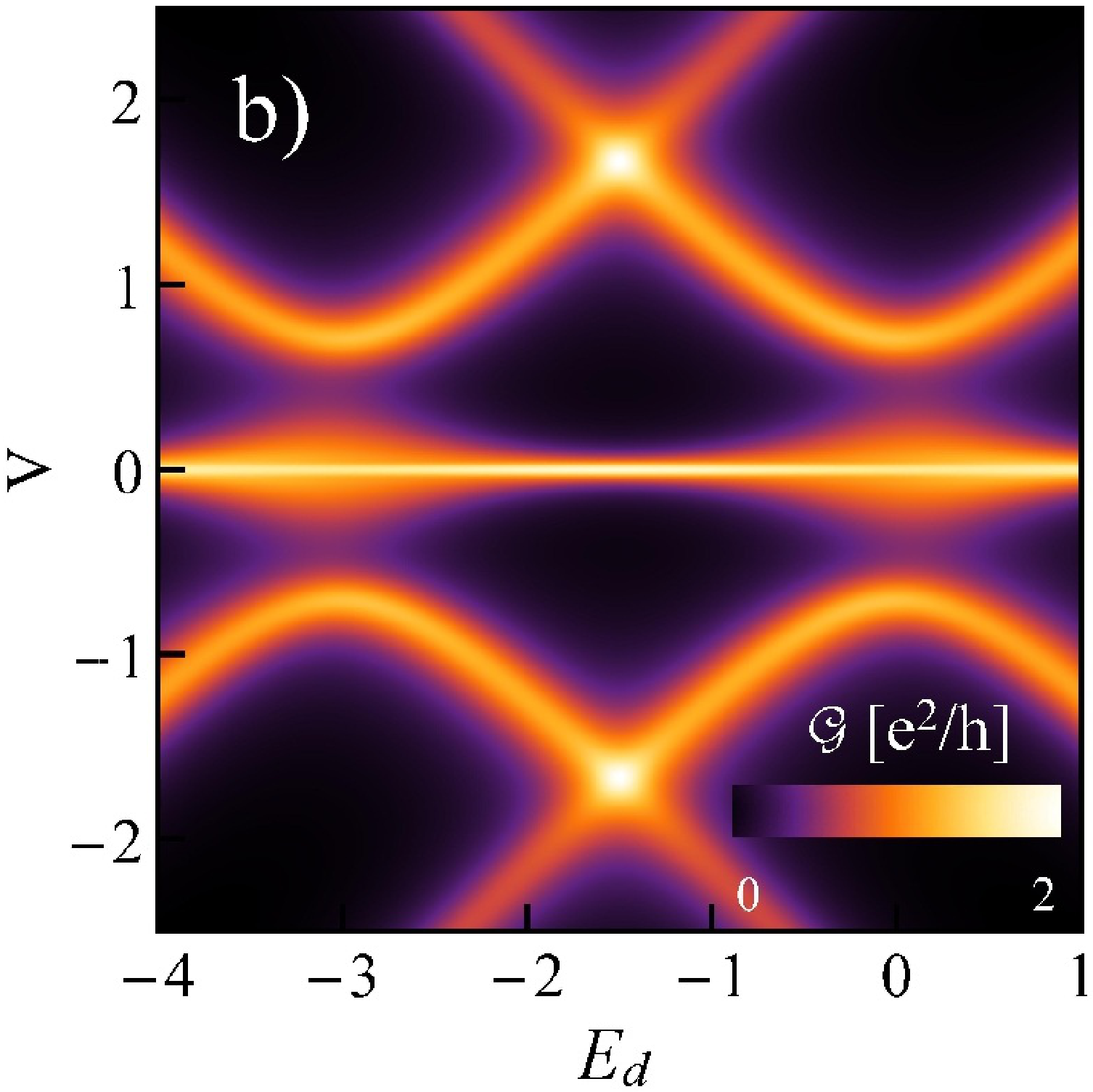}\\
\includegraphics[width=0.48\linewidth]{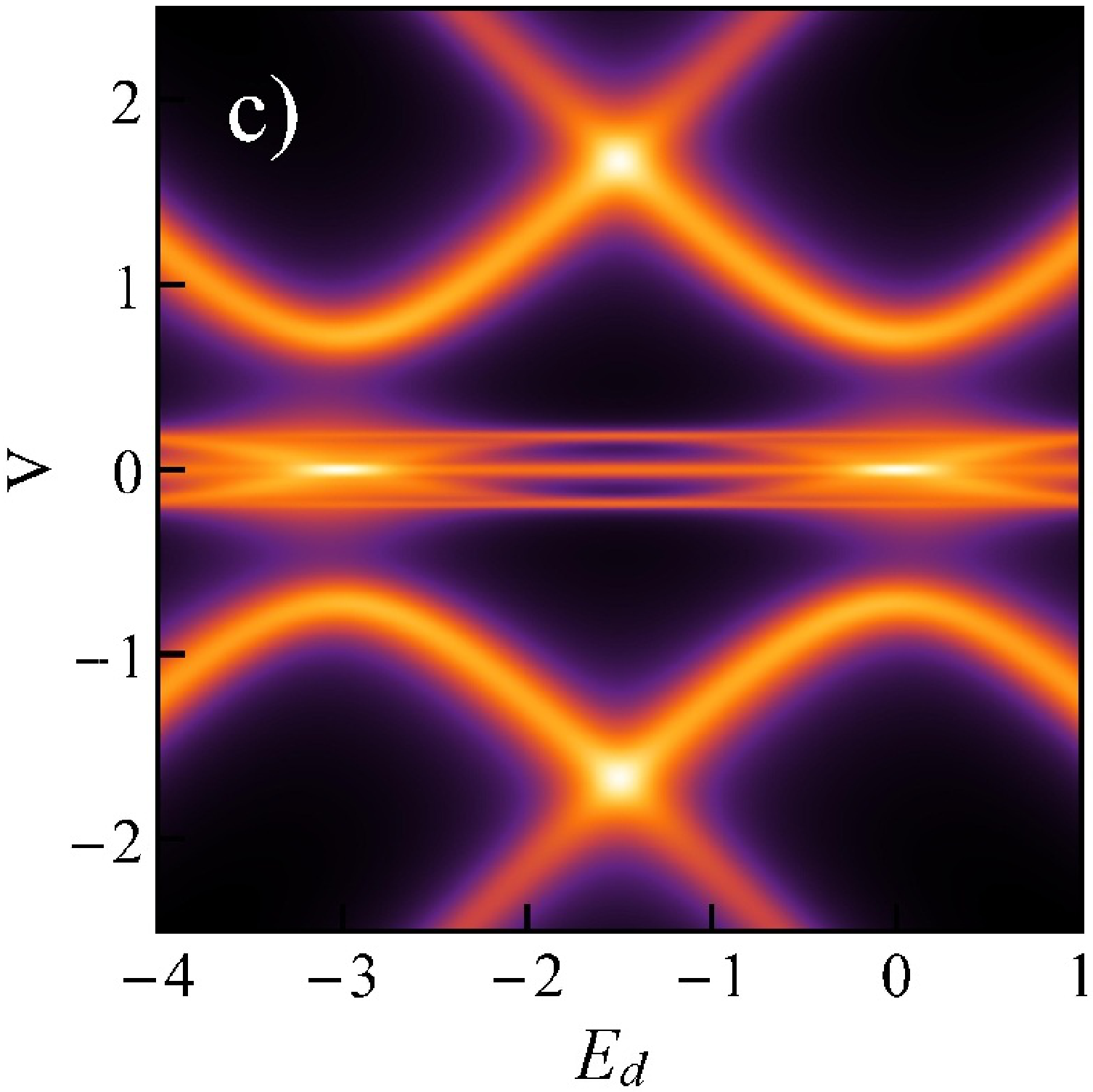}
\includegraphics[width=0.48\linewidth]{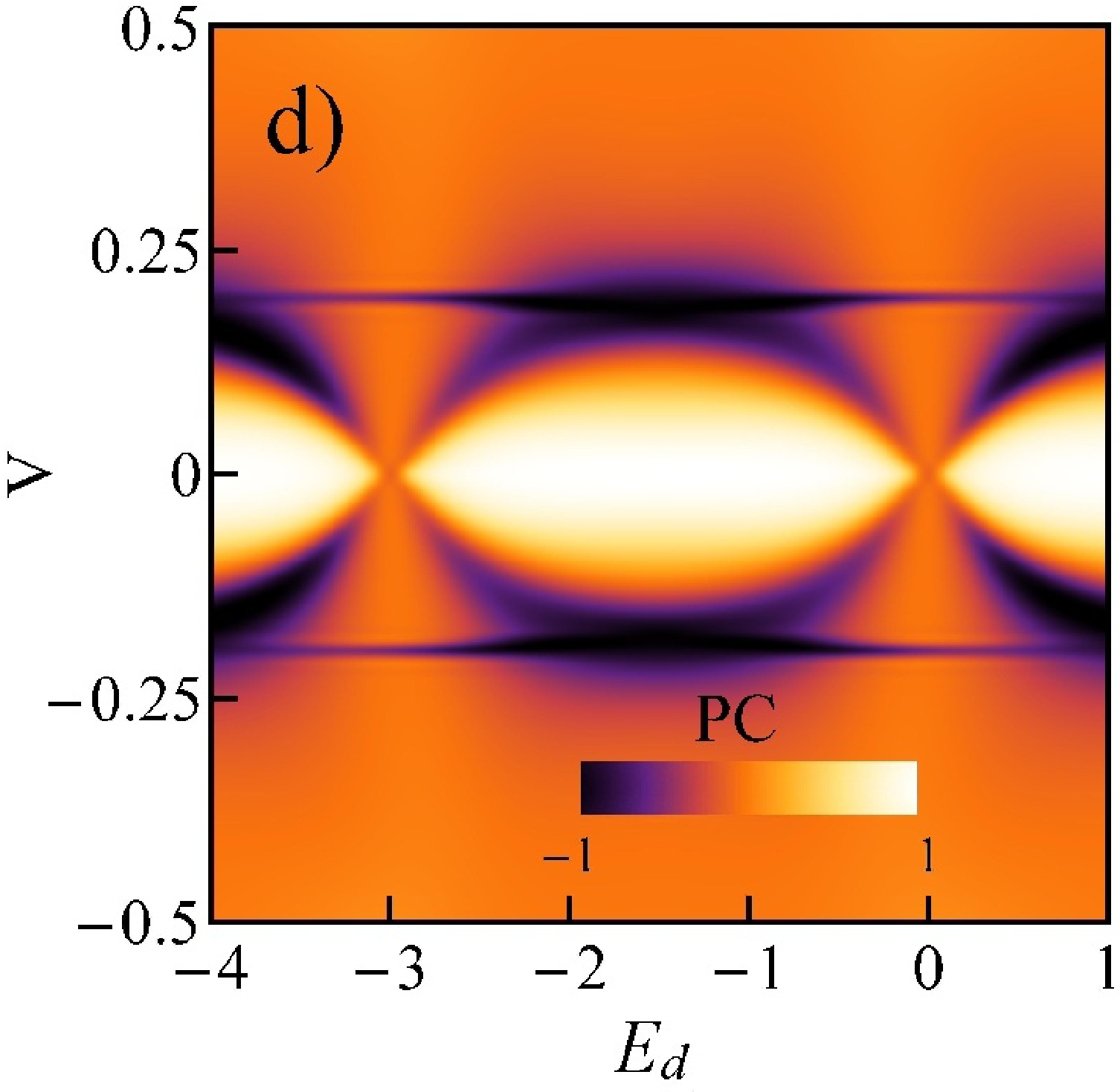}
\caption{\label{fig2} (Color online) a)   Polarizations of conductance for $E_{d} =-{\cal{U}}/2$ plotted vs. $t_{2}$ for several  fixed values of $t_{1}=t^{*}_{1}$  (right axis) or vs. $t_{1}$ for fixed values of $t_{2}=t^{*}_{2}$  (left axis), $t^{*}_{1(2)}=0.005$ (solid gray line), $t^{*}_{1(2)}=0.1$ (dotted blue), $t^{*}_{1(2)}=0.5$ (solid red) and $t^{*}_{1(2)}=1$ (dashed black). Lower inset presents a scheme of quantum dot coupled to two TS wires and upper inset shows total conductance map for $t_{1} = 0.5$ and $t_{2} = 0.25$ ($\Gamma=0.05$).   b,c) Differential conductance  for $\delta_{1} = \delta_{2} = 0$ (b) and $\delta_{1} =0$, $\delta_{2} = 0.2$. d) Polarization of conductance in the low bias region for $\delta_{1} =0$, $\delta_{2} = 0.2$ ($\Gamma = 0.25$, $ t_{1} = t_{1} = 0.5$).}
\end{figure}

\section{Results}
First we discuss the impact of single Majorana state on Coulomb blockade considering QD with only one side attached  TS wire ($t_{2}=0$) (inset of Fig. 1a).
Figs. 1c, d   show the spin  resolved differential conductance maps.  Zero bias anomaly (ZBA) with conductance reaching $(1/2)(e^{2}/h)$ is only observed for spin-up channel. Another signature  of Majorana-dot coupling is the observed splitting of Coulomb lines. Apart from typical Coulomb blockade diamonds also additional  lines with nonlinear bias and gate voltage dependencies appear. The former lines are   dominated by spin down electrons and the latter by tunneling of  spin up carriers.  Whereas spin up contribution is directly modified by Majorana-dot coupling, spin down  is influenced only indirectly by the change of occupation. The splitting between ordinary Coulomb blockade point and excited lines at two degeneracies points ($E_{d}=0$, $E_{d}=-{\cal{U}}$)  is of order of $t$.  Interesting observation is the occurrence of  spin dependent negative differential conductance  $d{\cal{I}}_{\sigma}/dV<0$ (SNDC) occurring for both spin orientations (blue regions at Figs. 1c, d). SNDC occurs on Coulomb blockade straight lines for spin up and  on the curved Coulomb lines for spin down.  To visualize this effect  more clearly we present also spin polarization maps, where in the regions of SNDC, according to the definition of polarization of conductance, $PC>+1$ occurs for spin up NDC (red regions on Figs. 1d, e) and $PC<-1$ for spin down (blue regions).
\begin{figure}
\includegraphics[width=0.45\linewidth]{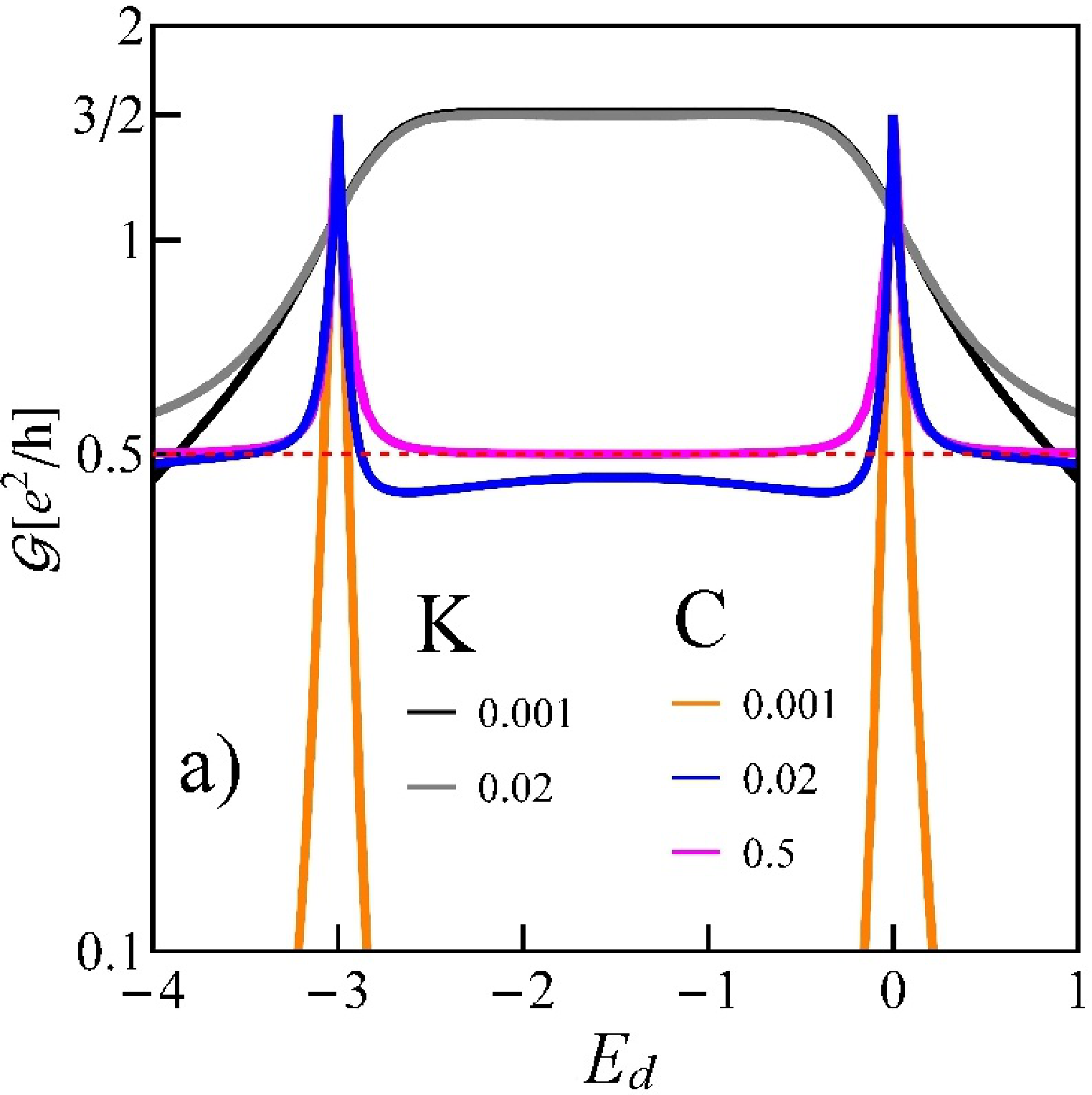}
\includegraphics[width=0.51\linewidth]{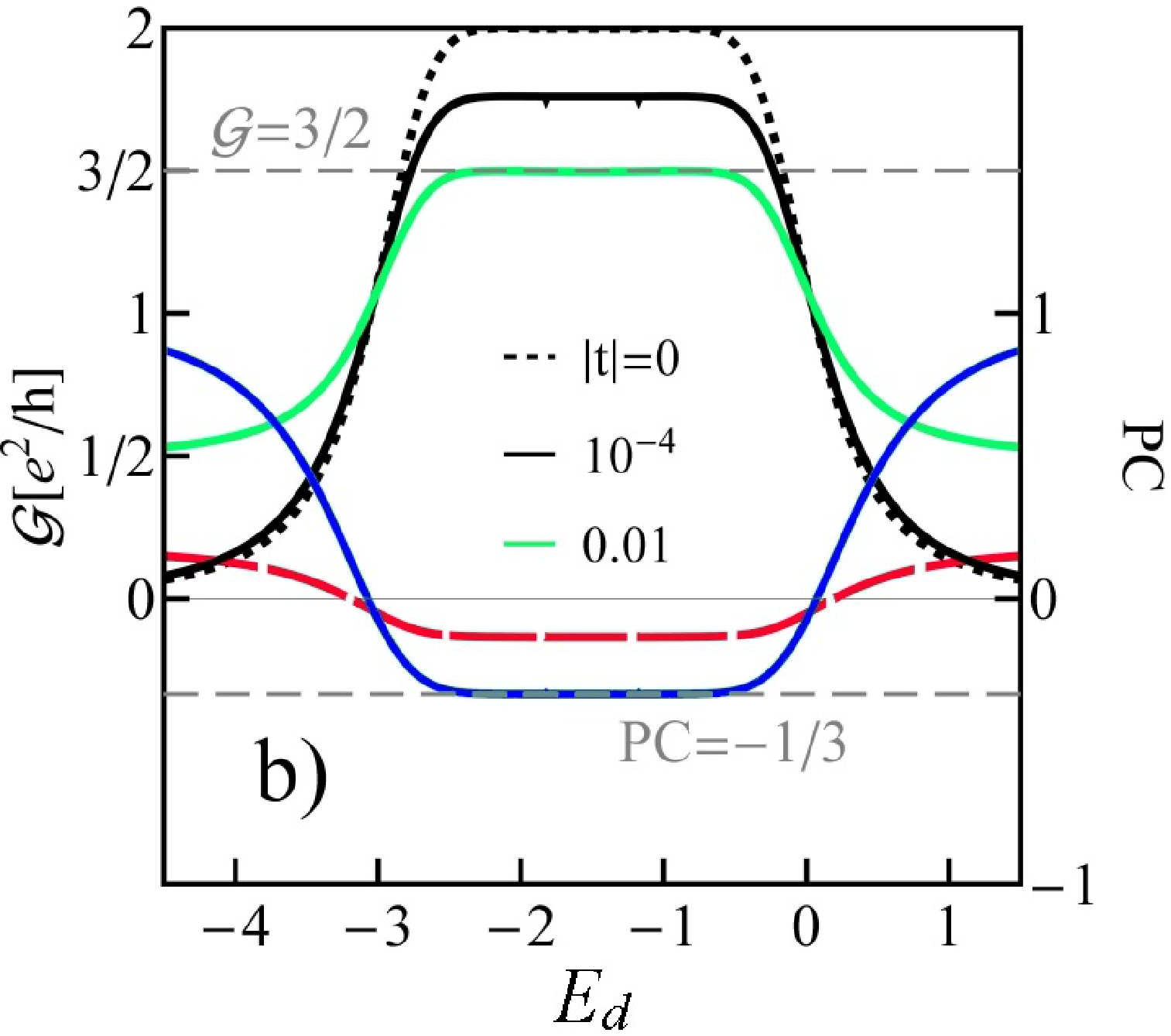}
\caption{\label{fig2} (Color online) a) Comparison of conductance  in the Kondo (K) and Coulomb blockade regimes (C) for different values of coupling of the dot with TS wire  t1 (t2 = 0). b) Conductances for the SU(2) Kondo (dotted black line, $t_{1}=0$), for SU(2) Majorana-Kondo effect (solid green, ${\cal{G}}=(3/2)$) and for intermediate case (black line). The solid blue and dashed red lines represent the polarization of conductance for $t_{1} = 10^{-4}$ and $t_{1} = 0,01$ (${\cal{U}} = 3$,  $\Gamma= 0,05$).}
\end{figure}
SNDC increases with weakening of the coupling to the normal electrodes.  Fig. 1f illustrates the effect of finite  overlap between Majorana fermions from the opposite ends of the wire ($\delta_{1}\neq0$).  ZBA resonance splits, but still remains  fully spin polarized.  Splitting is proportional to $\delta$.  Fig. 1b presents the  map of spin accumulation ($P=N_{\downarrow}-N_{\uparrow}$) with marked occupation numbers.
Note that in contrast to the unperturbed Coulomb blockade ($t=0$), for the system coupled to MBS ($t\neq0$) the additional intermediate occupations occur ($N=1/4, 3/4, 5/4, 7/4$) in the region between Majorana split Coulomb lines. In this range also spin accumulation changes the sign with decrease of $E_{d}$.
 Fig. 2 illustrates the impact of two Majorana modes of opposite spin polarizations.  Such a case can be realized   with two side attached   TS wires touching  the dot with the  opposite ends (Fig. 2a). For  $\delta_{1}=\delta_{2}=0$  and symmetric coupling ($t_{1}=t_{2}$)  conductance of  ZBA is doubled  in comparison to the case of coupling with  single wire  ($G=(e^{2}/h)$) and is unpolarized, because in tunneling equally engaged are carriers of both spin orientations.   For asymmetric coupling linear conductance becomes spin polarized.  Changing the tunneling amplitudes ($t_{1}/t_{2}\neq1$) one can modify the magnitude and the sign of polarization and these dependencies are sensitive to the strength of hopping parameters (Fig.2a).  When  Majorana modes from both wires  are directly coupled to the dot typical Coulomb diamonds disappear and only nonlinear conductance lines are visible (inset of Fig. 2b). Finite coupling  between two Majorana states in  the  wire leads to the earlier mentioned splitting of ZBA (Fig. 2c).  If  $\delta_{1}\neq\delta_{2}$  two pairs of finite bias lines appear characterized by opposite spin polarizations (not presented). Fig. 2d illustrates the special case ($\delta_{1}=0$ and  $\delta_{2}=0.2$), where both zero and finite bias lines coexist and the corresponding  polarizations of conductance are  opposite.  Changing the bias voltage one can switch between  the reversed polarizations (Fig. 2d).
Fig. 3 compares impact of ZBA on transport in the Coulomb range and Kondo regime and the main message is that due to the low energy scale of Kondo fluctuations only analysis of transport in this limit can  be used in the case of weak coupling of the dot to TS wire. As it is seen from Fig. 3a for $t = 0.001$ Coulomb blockade is undisturbed by the presence of MBS, contrary to Kondo case, where even for this small coupling  distinct reduction of Kondo conductance is observed. Fig 3b shows the evolution of Majorana perturbed Kondo conductance with the increase of $t$ and the spin polarization for $t\neq0$. The unitary conductance reaches $(3/2)(e^{2}/h)$ \cite{Lee} for $N=1$ and $(1/2)(e^{2}/h)$ for $N=0,2$. For SU(2) Majorana-Kondo effect, the polarization of conductance is negative for positively polarized Majorana state ($\gamma_{1}$)  ($PC=-1/3$), what is in contrast to the Coulomb range, where $PC=1$.

\begin{acknowledgments}
This project was supported by the Polish National Science Centre from the funds
awarded through the decision No. DEC-2013/10/M/ST3/00488.
\end{acknowledgments}

\def\refname{References}

\end{document}